\documentclass[conference]{IEEEtran}
\ifCLASSINFOpdf
\else
\fi
\usepackage{url}
\usepackage{amsmath}
\usepackage{algpseudocode}
\usepackage{algorithm}
\usepackage{hhline}
\usepackage{graphicx}
\usepackage{svg}
\usepackage{booktabs}
\usepackage{balance}

\begin{document}
%
\title{\textbf{AiR-ViBeR}: Exfiltrating Data from Air-Gapped Computers via Covert Surface ViBrAtIoNs.}

\author{\IEEEauthorblockN{Mordechai Guri}
\IEEEauthorblockA{Ben-Gurion University of the Negev, Israel\\Cyber-Security Research Center\\
gurim@post.bgu.ac.il \\ http://www.covertchannels.com \\ demo video: \textcolor{blue} {https://youtu.be/XGD343nq1dg}}}


%


\maketitle

\begin{abstract}
Air-gap covert channels are special types of covert communication channels that enable attackers to exfiltrate data from isolated, network-less computers. Various types of air-gap covert channels have been demonstrated over the years, including electromagnetic, magnetic, acoustic, optical, and thermal. 

In this paper, we introduce a new type of vibrational (seismic) covert channel. We observe that computers vibrate at a frequency correlated to the rotation speed of their internal fans. These inaudible vibrations affect the entire structure on which the computer is placed.  
Our method is based on malware's capability of controlling the vibrations generated by a computer, by regulating its internal fan speeds. We show that the malware-generated covert vibrations can be sensed by nearby smartphones via the integrated, sensitive \textit{accelerometers}. Notably, the accelerometer sensors in smartphones can be accessed by any app without requiring the user permissions, which make this attack highly evasive. We implemented AiR-ViBeR, malware that encodes binary information, and modulate it over a low frequency vibrational carrier. The data is then decoded by malicious application on a smartphone placed on the same surface (e.g., on a desk). We discuss the attack model, provide technical background, and present the implementation details and evaluation results. Our results show that using AiR-ViBeR, data can be exfiltrated from air-gapped computer to a nearby smartphone on the same table, or even an adjacent table, via vibrations. Finally, we propose a set of countermeasures for this new type of attack. 

\end{abstract}


%
\IEEEpeerreviewmaketitle

\section{Introduction}
Digital information processing, communication and storage are the backbone of modern
organizations. Information, however, seems to be the most
coveted asset of our era, and accordingly attracts malicious adversaries. Modern organizational networks are protected with a wide range of security products and monitoring systems including firewalls, intrusion prevention systems (IPSs), data leakage prevention (DLP) solutions, AV programs, and the like. When \textit{highly} sensitive data is involved, an organization may resort to air-gap isolation, in which there is no networking connection between the local network and the Internet. In this approach, any type connection between the internal network and the Internet is prohibited.

\subsection{Air-gap breaches}
Despite the level of isolation and lack of Internet connectivity, attackers have successfully compromised air-gapped networks in the past. Such networks can be breached by skillful combinations of devious malware and attack vectors \cite{Goodin2015,Karnouskos2011,Knowlton2010,DarkReading2006}. One of the most known incidents is the Stuxnet malware which penetrated uranium enrichment facility. In 2018, The US Department of Homeland Security accused Russian government hackers of penetrating America's power utilities \cite{Nobigdea65:online}. Due to reports in the Washington Post in November 2019, the Nuclear Power Corporation of India Limited (NPCIL) confirmed that the Kudankulam Nuclear Power Plant suffered a cyber-attack earlier that year \cite{AnIndian12:online}. 

\subsection{Air-gap exfiltration}
While breaching such systems has been shown feasible in recent years, exfiltration of data from systems without network or physical access is still considered a more challenging task. Most of the covert communication channels explored so far are electromagnetic  \cite{Kuhn1998,guri2014airhopper}, acoustic \cite{Hanspach2013,Halevi2012}, optical \cite{Loughry2002,guri2018xled} and thermal \cite{guri2015bitwhisper}. Accordingly, various countermeasures exists for these types of covert communication channels. 

\subsection{Vibration-based covert-channels}
In this paper, we introduce a new type of vibration-based (seismic) covert-channel. We show that malware can regulate the level of mechanical vibrations generated by a computer by controlling its fan rotation speed. These vibrations affect the entire structure on which computer is placed. Data can be encoded in these vibrations and then received by a smartphone placed on the table via the accelerometer sensor. We implement and evaluate the covert channel and test it in a typical workplace environment. 

\subsection{Accelerometers in smartphones}
The proposed air-gap covert channel is highly evasive from the attacker's perspective.
\begin{itemize}
	\item {\textbf{No permissions are needed.}} Smartphone accelerometers are considered safe sensors. Android and iOS operating systems applications do not request user permissions to read the outputs of the accelerometer samples. 
	\item {\textbf{No visual indication.}} There may be no visual indication to the user that an application is using the accelerometer. 
	\item {\textbf{JavaScript access.}} The accelerometer may be accessed from a Web browser via standard JavaScript code. This implies that the attacker Isn't required to compromise the user's device or install a malicious application. Instead, the attacker can implant  malicious JavaScript on a legitimate website that samples the accelerometer, receives the covert signals, and exfiltrates the information via the Internet.   
\end{itemize}

\subsection{Scope of this paper}
In this paper we show the feasibility of the proposed vibration-based covert channel, e.g., the capability of transmitting binary data over mechanical vibrations between workstations and smartphones. However, the physical layer of the vibrational covert channel is highly dependents on the specific environment, structure, and material of the surface and the location of the sender and receiver. Providing a comprehensive analytic model for the communication channel in a general case is beyond the scope of this paper and left for future work in the field of mechanical engineering.

\section{Related Work}
\label{sec:related}
Air-gap covert channels are special covert channels which enable communication from air-gapped computers, mainly for the purpose of data exfiltration. They are currently classified into five main categories: electromagnetic, magnetic, acoustic, thermal, and optical. This paper introduces a new category - the vibrational (seismic) covert channel.   

\subsection{Electromangetic}
Over the past 20 years, several studies have proposed the use of electromagnetic emanation from computers for covert communication. Kuhn  showed that it is possible to control the electromagnetic emissions from  computer displays \cite{kuhn1998soft}. Using this method, a malicious code can generated radio signals and modulate data on top of them. In 2014, Guri et al demonstrated AirHopper \cite{guri2014airhopper,guri2017bridging}, malware that exfiltrates data from air-gapped computers to a nearby smartphone via FM signals emitted from the screen cable. Later on Guri et al also demonstrated GSMem \cite{guri2015gsmem}, malware that leaks data from air-gapped computers to nearby mobile-phones using cellular frequencies generated from the buses which connect the RAM and the CPU. In 2016, Guri et al presented USBee, a malware that uses the USB data buses to generate electromagnetic signals from a desktop computer \cite{guri2016usbee}. In 2018, Guri et al also presented PowerHammer,
an attack vector for leaking data from air-gapped computers through the power lines \cite{guri2019powerhammer}.
 
\subsection{Magnetic}
In 2018, Guri et al presented ODINI \cite{guri2019odini}, malware that can exfiltrate data from air-gapped computers via low frequency magnetic signals generated by the computer's CPU cores. The magnetic fields bypass Faraday cages and metal shields. Guri et al also demonstrated MAGNETO \cite{guri2018magneto}, which is malware that leaks data from air-gapped computers to nearby smartphones via magnetic signals; they used the magnetic sensor integrated in smartphones to receive covert signals. 

\subsection{Optical}
Several studies have proposed the use of optical emanation from computers for covert communication. Loughry introduced the use of PC keyboard LEDs to encode binary data \cite{loughry2002information}. In 2019 research explored the threat of keyboard LEDs with modern USB keyboards \cite{guri2019ctrl}.
In 2017, Guri et al presented LED-it-GO, a covert channel that uses the hard drive indicator LED in order to exfiltrate data from air-gapped computers \cite{Guri2017}. Guri et al also presented a method for data exfiltration from air-gapped networks via routers and switch LEDs \cite{guri2018xled}. Data can also be leaked optically through fast blinking images or low contrast bitmaps projected on the LCD screen \cite{guri2016optical}. In 2017, Guri et al presented aIR-Jumper, malware that uses the security cameras and their IR LEDs to communicate with air-gapped networks remotely \cite{guri2019air}. In 2019 researchers introduced a covert channel (dubbed BRIGHTNESS) which uses the LCD  brightness to covertly modulate information and exfiltrate it to a remote camera \cite{guri2019brightness}. 

\subsection{Thermal}
In 2015, Guri et al introduced  BitWhisper \cite{guri2015bitwhisper}, a thermal covert channel allowing an attacker to establish bidirectional communication between two adjacent air-gapped computers via temperature changes. The heat is generated by the CPU/GPU of a standard computer and received by temperature sensors that are integrated into the motherboard of the nearby computer. Unlike BitWhisper which works between two adjacent desktop computers, this paper discuss the thermal covert channel between a desktop workstation and a nearby mobile phone. 

\subsection{Acoustic}
In acoustic covert channels, data is transmitted via inaudible, ultrasonic sound waves. Audio based communication between computers was reviewed by Madhavapeddy et al. in 2005 \cite{madhavapeddy2005audio}. In 2013, Hanspach \cite{hanspach2014covert} used inaudible sound to establish a covert channel between air-gapped laptops equipped with speakers and microphones. Their botnet established communication between two computers located ~19 meters apart and can achieve a bit rate of 20 bit/sec. Deshotels \cite{deshotels2014inaudible} demonstrated the acoustic covert channel with smartphones, and showed that data can be transferred up to 30 meters away. In 2013, security researchers claimed to find malware (dubbed BadBios) that communicates between two instances of air-gapped laptops via the integrated speakers and microphones using ultrasonic signals \cite{Meetbad27:online}. 
\\

All of the acoustic methods presented above require speakers.  In 2016, Guri et al introduced Fansmitter, a malware which facilitates the exfiltration of data from an air-gapped computer via noise intentionally emitted from the PC fans \cite{guri2016fansmitter}. In this method, the transmitting computer does not need to be equipped with audio hardware or an internal or external speaker. Guri et al also presented DiskFiltration a method that uses the acoustic signals emitted from the hard disk drive's moving arm to exfiltrate data from air-gapped computers \cite{guri2017acoustic}. Guri et al presented Speake(a)r \cite{guri17speake} malware that covertly turns the headphones, earphones, or simple earbuds connected to a PC into a pair of eavesdropping microphones when a standard microphone is muted, taped,  turned off, or not present.

\begin{table}[]
	\centering
	\caption{Summary of existing air-gap covert channels}
	\label{table-related}
	\begin{tabular}{ll}
		\hline
		Type               & Method                                                                                                                                                                                          \\ \hhline{==}
		Electromagnetic    & \begin{tabular}[c]{@{}l@{}}AirHopper \cite{guri2014airhopper,guri2017bridging} (FM radio) \\ GSMem \cite{guri2015gsmem} (cellular frequencies)  \\ USBee \cite{guri2016usbee} (USB bus emission) \\ Funthenna \cite{funtenna86:online} (GPIO emission) \\
		PowerHammer \cite{guri2019powerhammer} (power lines)	
	 \end{tabular}                                                                   \\ \hline
		Magnetic           & \begin{tabular}[c]{@{}l@{}}MAGNETO \cite{guri2018magneto} (CPU-generated \\ magnetic fields)\\ ODINI \cite{guri2019odini} (Faraday shields bypass) \\ Hard-disk-drive \cite{matyunin2016covert} \end{tabular}                                                                                            \\ \hline
		Acoustic           & \begin{tabular}[c]{@{}l@{}}Fansmitter \cite{guri2020fansmitter,guri2016fansmitter} (computer fan noise) \\ DiskFiltration \cite{guri2017acoustic} (hard disk noise) \\ Ultrasonic \cite{hanspach2014covert,carrara2014acoustic}\\ MOSQUITO (speaker-to-speaker)\end{tabular} \\ \hline \\
		Thermal            & \begin{tabular}[c]{@{}l@{}} BitWhisper  \cite{guri2015bitwhisper} 
		\end{tabular}                                                                                                                                                                             \\\\ \hline 
		Optical            & \begin{tabular}[c]{@{}l@{}}LED-it-GO \cite{Guri2017} (hard drive LED) \\ VisiSploit \cite{guri2016optical} (invisible pixels) \\ Keyboard LEDs \cite{loughry2002information}  \cite{guri2019ctrl}\\ Router LEDs \cite{guri2018xled} \\
		aIR-Jumper \cite{guri2019air} (security cameras)  \\
		BRIGHTNESS (LCD brightness) \cite{guri2019brightness} 
	\end{tabular}                            \\ \hline \\

		 Vibration (Seismic) &
		\begin{tabular}[c]{@{}l@{}} AiR-ViBeR, this paper (computer vibrations) \end{tabular}                                                            \\ \hline
	\end{tabular}
\end{table}
Table \ref{table-related}. summarizes the existing air-gap covert channels.

\section{Attack Model}
The adversarial attack model consists of a transmitter and a receiver. In this scenario, the transmitter is a desktop workstation, and the receiver is a nearby smartphone belonging to an employee. 

\begin{figure}
	\centering
	\includegraphics[width=1.0\linewidth]{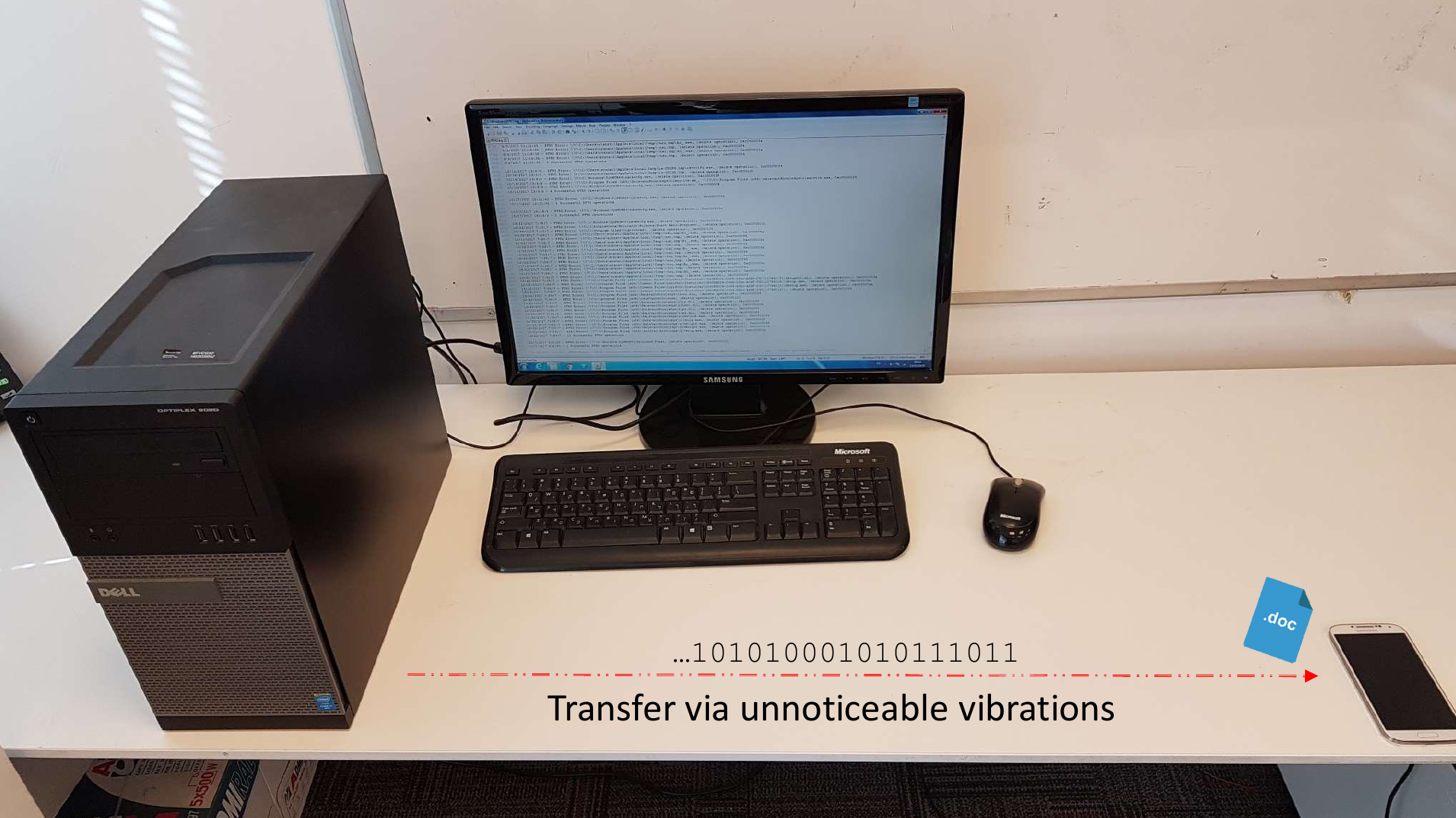}
	\caption{Illustration of the cover channel. The malware in the compromised computer transmits signals to the environment via vibrations induced on the table. A nearby infected smartphone detects the transmission, demodulates and decodes the data, and transfers it to the attacker via the Internet.}
	\label{fig:illustration}
\end{figure}

\
In a first stage of the attack, the transmitter and receiver are compromised by the attacker. Infecting highly secure networks can be accomplished, as demonstrated by the attacks involving Stuxnet \cite {langner2011stuxnet}, Agent.BTZ \cite{grant2009cyber}, and others \cite{TheEpicT20:online,zaored,AFannyEq68:online}. In addition, mobile phones of employees are identified, possibly by using social engineering techniques. The employees are assumed to carry their mobile phones around the workplace. These devices are then infected either online, using a device's vulnerabilities, or by physical contact if possible. Infecting a mobile phone can be accomplished via different attack vectors, using emails, SMS/MMS, malicious apps, malicious websites, and so on \cite{provos2007ghost, cova2010detection, sood2011malvertising, peltier2006social, smutz2012malicious}.  After gaining a foothold in the organization, malware in the compromised computer gathers the information of interests (e.g., encryption keys, key-logging, etc.).
In the exfiltration phase, the malware encodes the data and transmits it to the environment via vibrations on the surface (Figure \ref{fig:illustration}). A nearby infected smartphone detects the transmission with its accelerometer, demodulates and decodes the data, and then transfers it to the attacker via the Internet.

\subsection{Accelerometer permissions}
Note that the adversary can install applications on the mobile device without requiring special permissions. 
Smartphone accelerometers are considered safe sensors, and mobile OS (Android and iOS) do not request user permissions to read the outputs of the accelerometer samples. 
Furthermore, the accelerometer may be accessed from a Web browser via  standard JavaScript code. This implies that the attacker doesn't need to compromise the user's device via a malicious application. Instead, the attacker can implant malicious JavaScript on legitimate websites that sample the accelerometer's data.

\section{Technical Background}
\subsection{Computer fans}
\label{sec:tech}
Various computer components such as the CPU, GPU, RAM and HDD produce heat during normal usage. These components must be kept within a specified temperature range in order to prevent overheating, malfunction, and damage. Computer fans accelerate the cooling of these components by increasing air flow around them.	
Desktop computers are typically equipped with three to four fans as listed below. 

\begin{enumerate}
	\item \textbf{The PSU (power supply unit) fan.} This fan is integrated into the PSU at the back of the unit. It is used as an exhaust fan to expel warm air from the PSU which produces heat. This type of fan is managed by an internal controller and usually cannot be monitored, controlled, or regulated by software.
	
	\item \textbf{The chassis fan.} This fan is installed on the side or at the back of the computer case. It usually draws in cold air from outside the computer and expels it through the top or rear of the computer.  
	
	\item \textbf{The CPU fan.} This fan is mounted on top of the CPU socket. It cools the CPU's heatsink.
	
	\item \textbf{The GPU fan}. Due to the large amount of heat emitted from modern graphics cards, they are shipped with dedicated cooling fans. Like CPU fans, they are mounted to the heatsink of the GPU. Some GPUs are shipped with a single fan, while others have multiple fans.
\end{enumerate}

\begin{figure}
	\centering
	\includegraphics[width=0.7\linewidth]{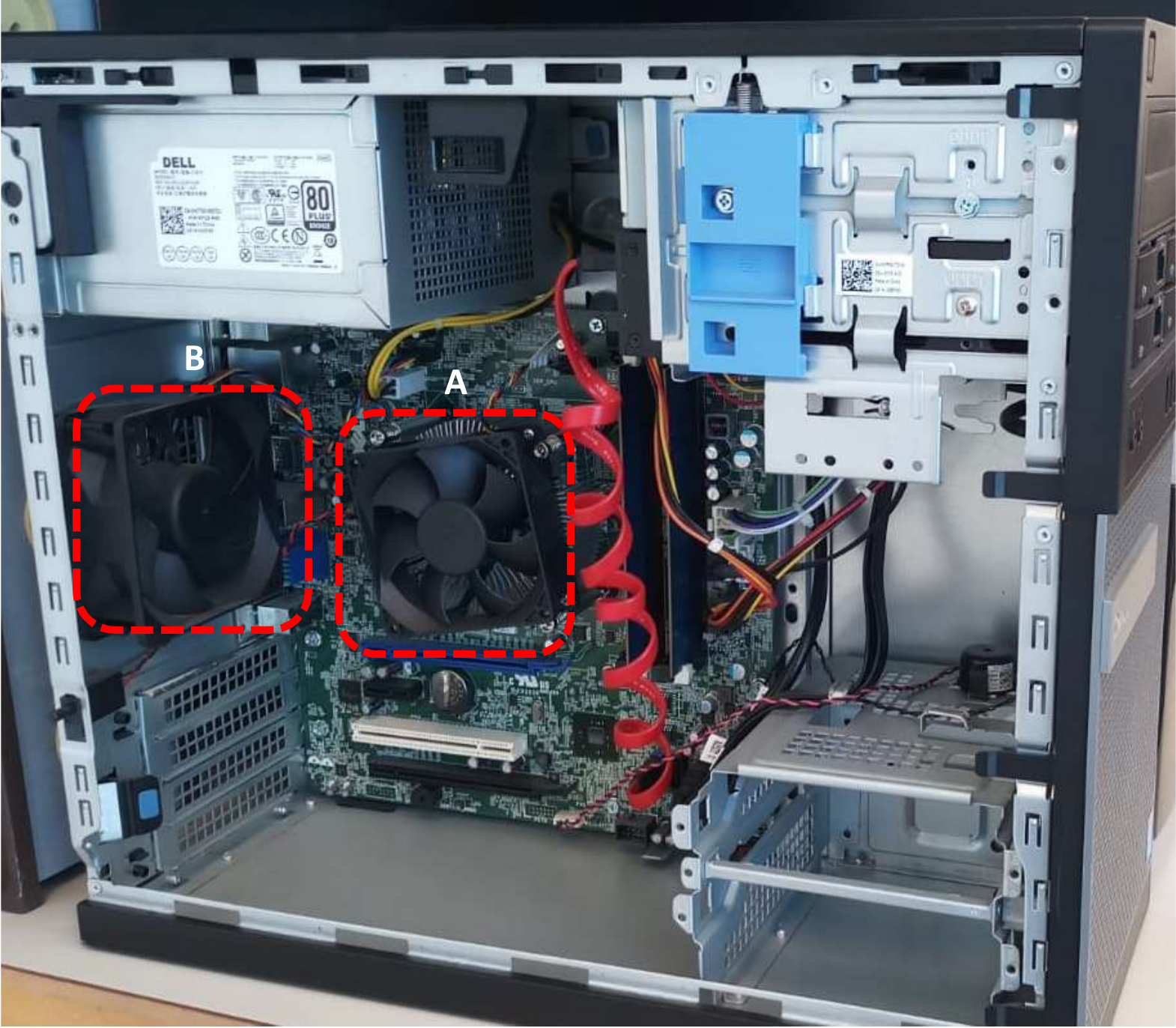}
	\caption{The CPU (A) and chassis (B) fans within a typical workstation.}
	\label{fig:case}
\end{figure}

There are other types of computer fans which are less common in desktop workstations. These include HDD fans, PCI fans, and memory slot fans, which might be found in server systems or legacy equipment. Table \ref{tab:fanlist} lists the computer fans and basic characteristics discussed above. Figure \ref{fig:case} shows the CPU  and chassis fans within a typical workstation.

\begin{table}[]
	\centering
	\caption{Computer fans}
	\label{tab:fanlist}
	
	\begin{tabular}{@{}lll@{}}
		\toprule
		Fan                                                                     & Availability & Speed control \\ \midrule
		PSU fan                                                                 & High        & No           \\
		Chassis fan                                                             & High        & Yes          \\
		CPU fan                                                                 & High        & Yes          \\
		GPU fan(s)                                                              & Medium      & Yes          \\
		\begin{tabular}[c]{@{}l@{}}Others (e.g., HDD, RAM, PCI)\end{tabular} & Low         & Yes          \\ \bottomrule
	\end{tabular}
\end{table}

In this paper we mainly focus on the chassis fans which generate the highest level of vibrations. This fan is present in all computers, making AiR-ViBeR a threat to virtually almost any desktop computer today. The PSU fan has been omitted from our discussion, since in most cases it can not be controlled via software. 

\subsection{Fans rotation}
Computer fan rotation, measured in revolutions per minute (RPM) units, emits acoustic noise at various frequencies and strengths. Typical computer fan speeds range from a few hundred to a few thousand RPMs. The movement of the fan blades, each of which pushes air in its path, creates a compression wave with some amount of rarefaction. The vibration level depends on the air flow, mechanics, location, size, number of blades, and current RPM of the rotating fan. Because the location, size, and number of blades of a fan are fixed, the current RPM is the main factor that contributes to variations in the vibrations. Given a fan rotating at $R$ RPM, the vibration induced on the surface will be at a frequency of $R/60$ Hz.

\subsection{Rotating Unbalance}
The vibrations caused by rotors and fans have been  studied extensively in the literature related to mechanical engineering. In this section we describe the basic concept of \textit{rotating unbalance}. We refer the interested reader to related literature on this topic \cite{article,czmochowski2014tests,buzdugan1986examples}.

Unbalance is defined as an uneven distribution of mass around an axis, causing the mass axis to differ from the bearing axis. When an object is rotating, the unequal mass along with the radial acceleration create a centrifugal force. This results in force on the bearings which causes the bearing to vibrate. Note that even a small amount of unequal mass can produce a noticeable amount of rotating unbalance force.

\begin{figure}
	\centering
	
	\includegraphics[width=0.5\linewidth]{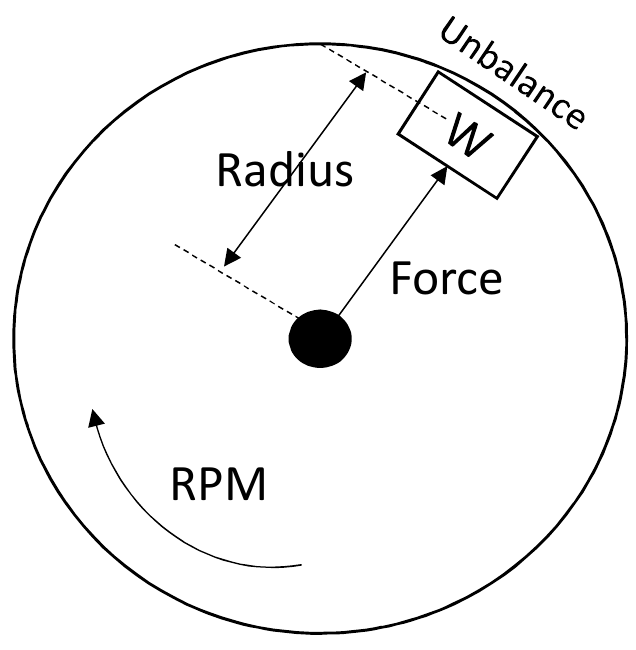}
	\caption{Rotating unbalance}
	\label{fig:unbalance}
\end{figure}


In Figure \ref{fig:unbalance}, a source of unbalance is shown as an object $W$, located at distance radius $R$, from the rotating centerline. If the unbalance weight, radius and  motor rotation speed ($RPM$) are known, the centrifugal force, $F$, can be calculated:

\begin{equation}
t = \frac{1}{RPM}
\end{equation}

\begin{equation}
V = \frac{2×\pi ×r}{t}
\end{equation}

\begin{equation}
Velocity(V) =  \frac{Circumference(C)}{Time(t)}
\end{equation}

\begin{equation}
F = \frac{W × (\frac{2×\pi ×r}{t})^2}{R}
\end{equation}

\begin{equation}
Centrifugal Force(F) = Mass(W) × V^2
\end{equation}

For example, for $W=0.1 \: ounce$, $R=2 \: inches$ and $RPM=2600$ the amount of centrifugal force produced is 
  
\begin{equation}
F=(0.1 \: ounce)*\frac{(\frac{\frac{2*\pi*3 \: inches}{1}}{2600 \: RPM})^2}{3 \: inches}
\end{equation}

The unbalance of computer fans can stem from many factors, including geometrical eccentricity, difference in the shape of the blades, corrosion, manufactured unsymmetrical configurations and more \cite{Unbalanc31:online}. The most common cause of unbalance in fans is the accumulation of material or the wear of the fan blades, which depend on the fan's operation. All of these situations cause a radial unbalance of the fan's mass \cite{Diagnosi99:online}. It is possible that with time, a fan may have more sources of unbalance and hence may generate increased vibrations over time.

\subsection{Smartphone accelerometers}
An accelerometer is an electromechanical device used to measure acceleration forces. Acceleration is the measurement of the change in velocity, or speed divided by time. 
All modern smartphones have an integrated accelerometer. This sensor is utilized by a wide range of applications such as device pairing, navigation, activity recognition and so on. The smartphone accelerometer measures the acceleration of the device on the $x$, $y$, and $z$ axes. The measurement of acceleration is provided in units of meters per second squared:

\begin{equation}
acc = m / sec^2
\end{equation}

Typical smartphone accelerometers are capable of measuring the acceleration in the $x$, $y$, and $z$ planes to a precision of six decimal places. For example, the Samsung Galaxy S10 used in our experiments has an integrated LSM6DO accelerometer which has a resolution of  0.0023956299  $m/sec^2$.

\subsection{Covert channel}
Because of their high sensitivity, accelerometers can expose privacy information to attackers. For example, TouchLogger \cite{cai2011touchlogger}, TapLogger \cite{xu2012taplogger} and ACCessory \cite{owusu2012accessory} demonstrate how attackers can recover keystrokes on touch screens from smartphone motions. In this paper, we propose using the accelerometer to measure covert vibrations: malware within the computer generates vibrations by changing the workstation fan speed; these vibrations are induced on the surface (e.g., a table) and measured by a malicious application within the smartphone using the accelerometer. 

\begin{figure}
	\centering
	\includegraphics[width=1.0\linewidth]{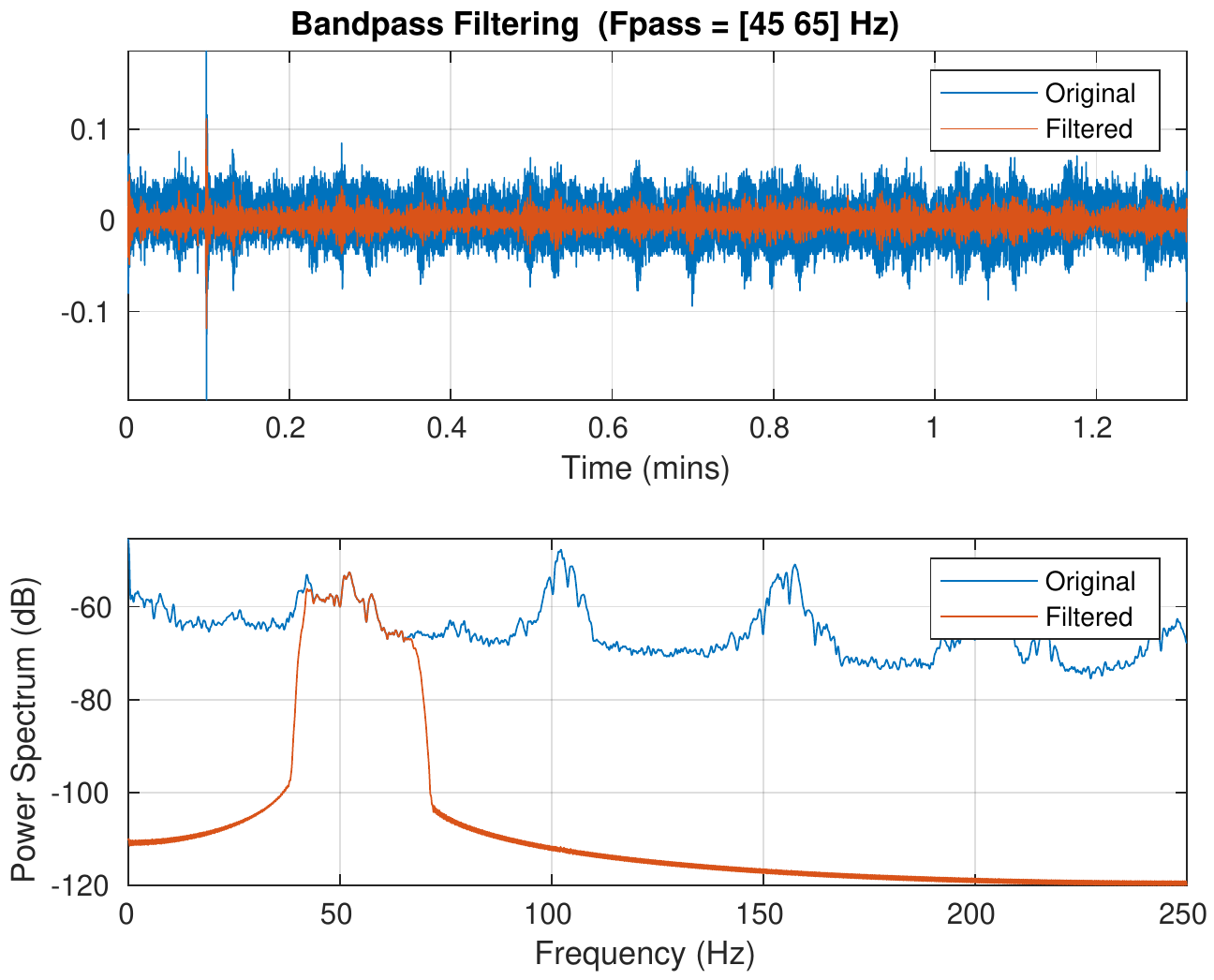}
	\caption{The raw vibration signal (bandpass filtered) generated by fan rotation at 3000RPM  and the power spectral density (PSD) graph. The vibrations are measured 70cm from the desktop computer.}
	\label{fig:POWERSPECTRUM}
\end{figure}

\begin{figure}
	\centering
	\includegraphics[width=0.5\linewidth]{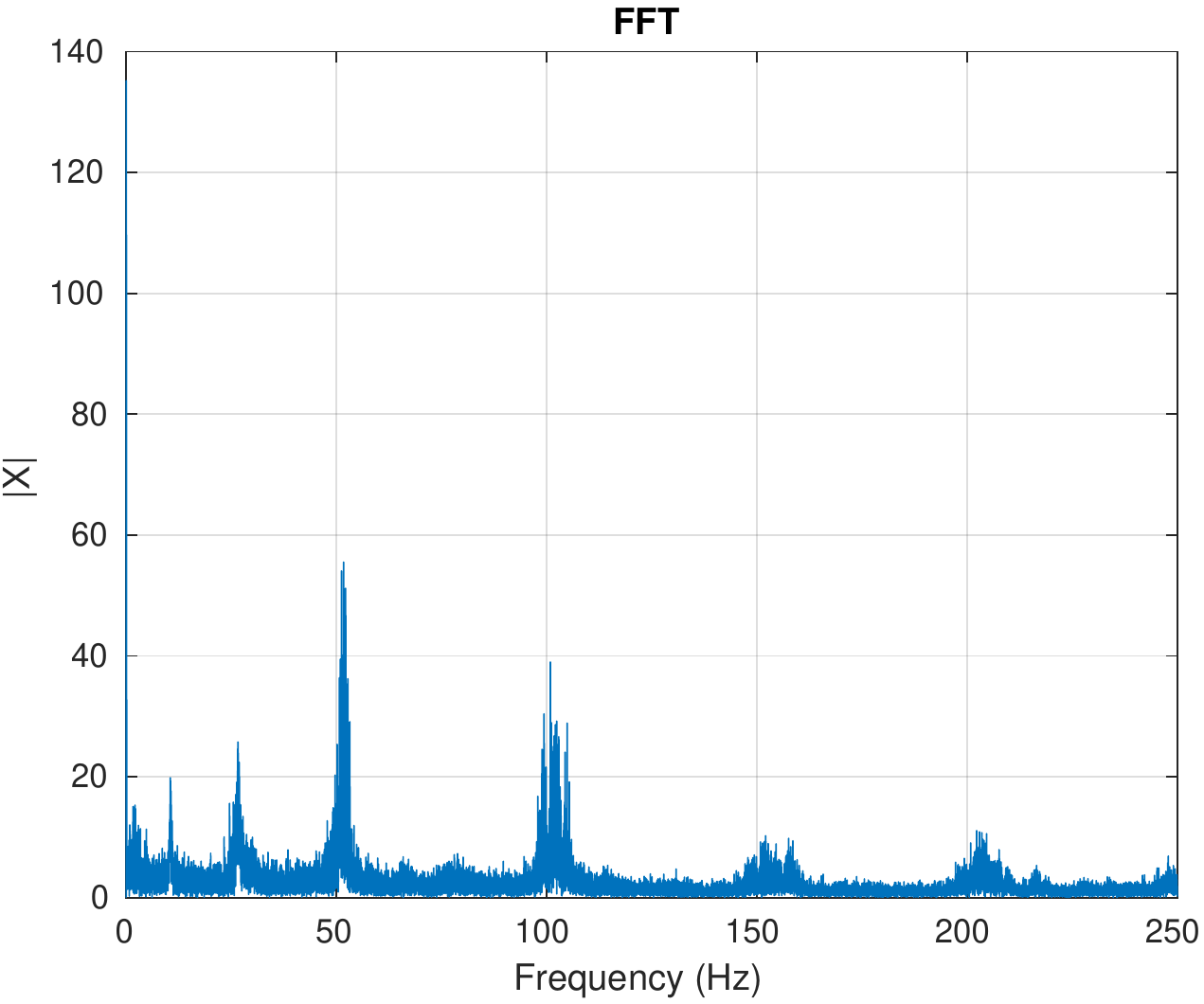}
	\caption{The FFT of the vibration signal generated by fan rotation at 3000RPM. The vibrations are measured 70cm from the desktop computer.}
	\label{fig:FFT}
\end{figure}

\section{Implementation}
\label{sec:imp}
In this section we describe signal generation and present the data modulation schemes and transmission protocol.

\begin{figure}[t]
	\centering
	\includegraphics[width=1.0\linewidth]{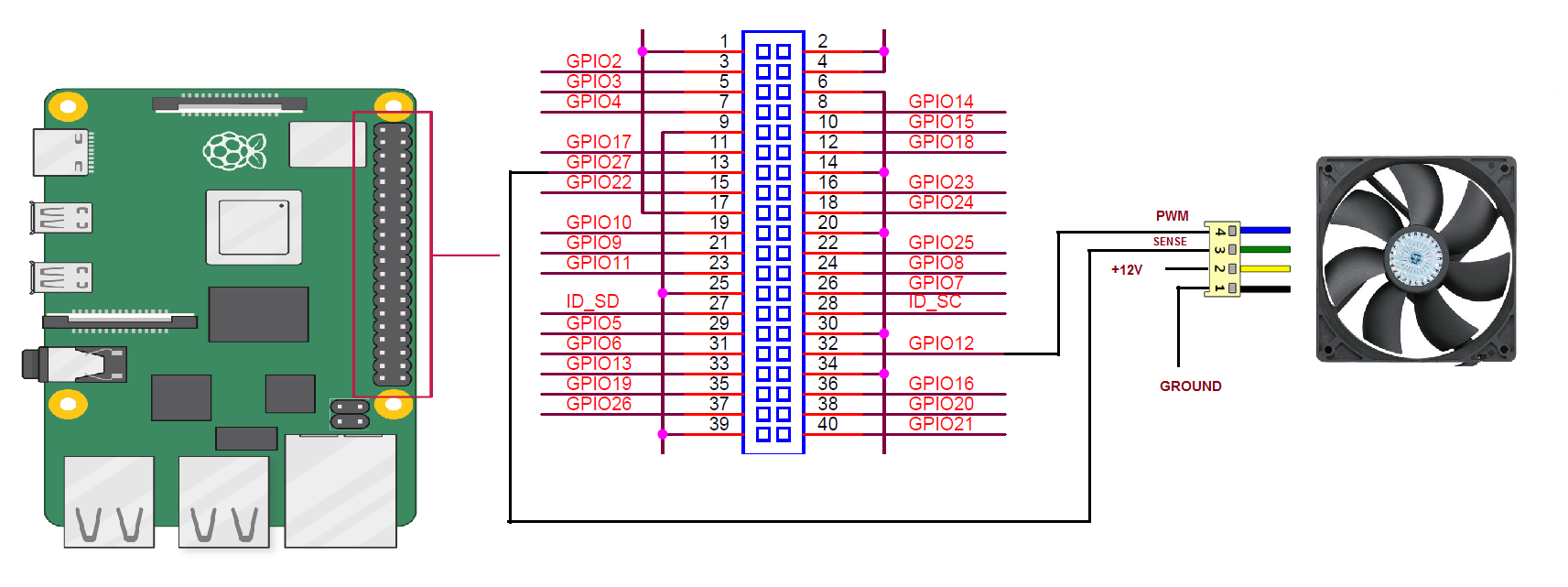}
	\caption{The chassis fan control setup. Pin 4 (fan control pin) of the fan is attached to a Raspberry Pi 3 GPIO12 (PWM GPIO).}
	\label{fig:RPI}
\end{figure}

\subsection{Fan control}
CPU and chassis fans are connected to pin headers on the motherboard via three or four wire connectors. Pins 1-2 are the ground and 12V power pin, respectively.  
Pin 3 (FAN\_TACH) is used for input, allowing the controller on the motherboard to sample the current RPM. Pin 4 (FAN\_CONTROL) is used for output, allowing the motherboard to control the fan speed via a pulse-width modulation signal (PWM) \cite{SpeedFan80:online}. In order to control the chassis fan speed from a standard Python script, we attached the FAN\_CONTROL pin to a Raspberry Pi 3 GPIO12 pin. This setup allows the initiation of PWM output directly to the fan controller by our programs (Figure \ref{fig:RPI}). During the experiments the Raspberry Pi was located within the computer chassis.

\subsection{Signal generation}
We generated vibrations by manipulating the rotation speed of the chassis fan. Our experiments show that there is a direct correlation between the rotation speed of the fan and the vibrations induced on the surface (due to the rotating unbalance). Given a fan rotating at $R$ RPM, the vibration induced on the surface will be at a frequency of $R/60$ Hz. Figures \ref{fig:POWERSPECTRUM} and \ref{fig:FFT} show the power spectral density (PSD) and the FFT graphs of the vibration signal generated by a chassis fan rotating at 3000 RPM, as recorded by a nearby smartphone on the table. The vibrations at $3000/60=50$ Hz can be seen in the graphs. 

\subsection{Modulation}
We present two modulation schemes based on amplitude shift keying (ASK) and frequency shift keying (FSK).

\begin{algorithm} 
	\caption{modulateRTZ (fan, $RPM_{0}$, $RPM_{base}$, $RPM_{1}$, bits, stateDuration, bitDuration)} 
	\begin{algorithmic}[1] 

		\State $ bitStart = getCurrentTime() $
		\For {$ bit\ in\ bits $}
		\If { $ bit == 1 $ }
		\State $ fan.setRPM(RPM_{1}) $
		\Else
		\State $ fan.setRPM(RPM_{0}) $
		\EndIf
		\\
		\State $ sleep(bitStart + stateDuration - getCurrentTime())$
		\State $ fan.setRPM(RPM_{base}) $
		\\
		\State $ bitStart\ +=\ bitDuration $
		\State $ sleep(bitStart - getCurrentTime())$
		\EndFor
		
	\end{algorithmic}
	\label{alg:a1}	 
\end{algorithm}

\subsection{Frequency-shift keying (FSK)}
\begin{figure}
	\centering

	\includegraphics[width=0.8\linewidth]{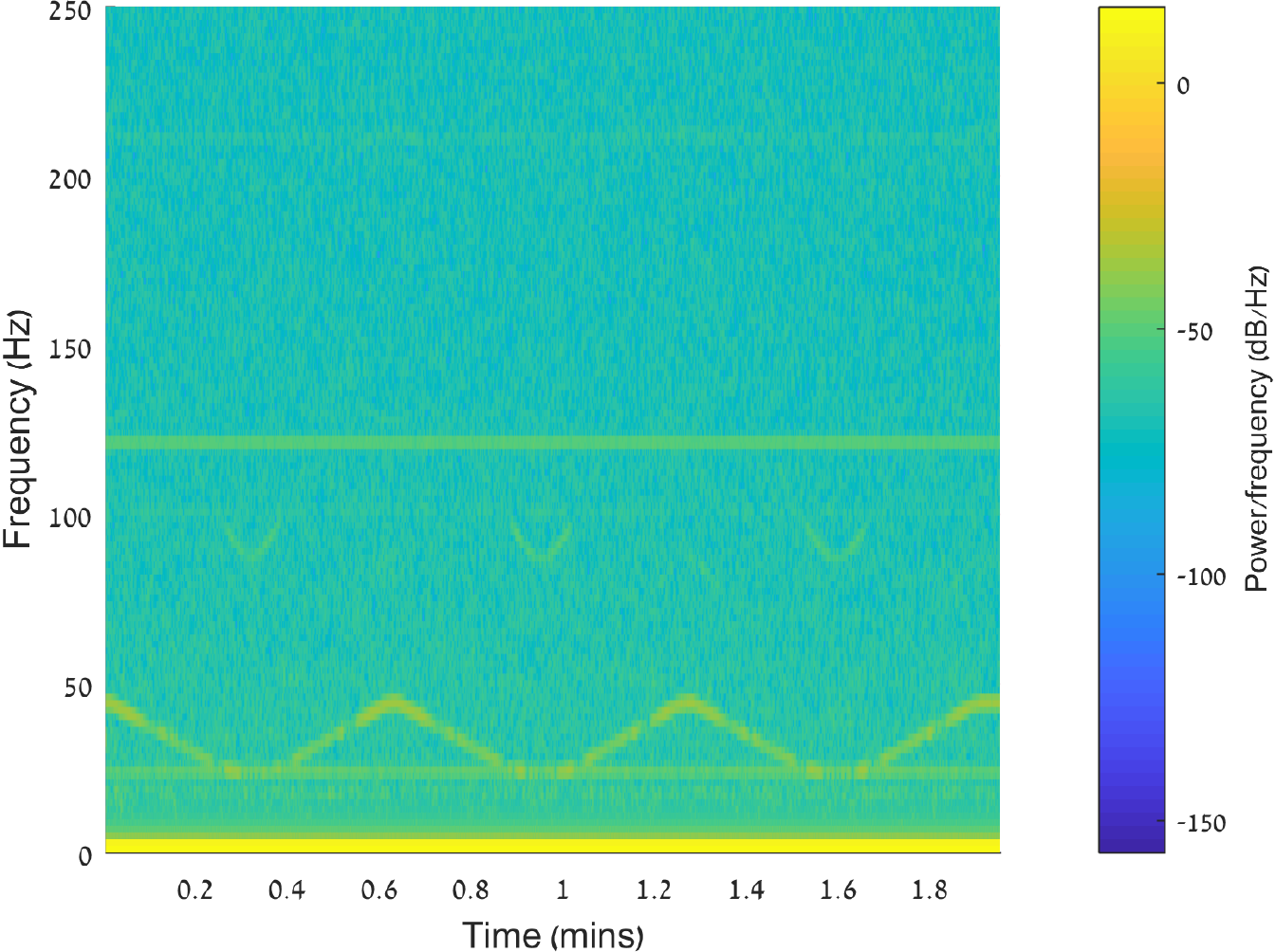}
	\caption{Spectrogram of an FSK modulation. The vibration generated with $RPM_{0}$ = 1300 and $RPM_{1}$ = 2600. In this case the sequence '10101010' has been transmitted.}
	\label{fig:FSK1}
\end{figure}

In the frequency-shift keying modulation, we assign distinct frequencies to represent distinct values of binary data. We use a version of FSK in which two distinct frequencies, $f_{0}$  and $f_{1}$, represent '0' and '1' arbitrarily. A third frequency ($RPM_{base}$) is used to separate between sequential bits. As mentioned, the vibration frequency is determined by the current RPM, and a change to the RPM results in a change in the vibrational frequency such as $f=RPM/60$. We maintain the frequency of the carrier by setting the fan to rotate at two speeds, $RPM_{0}$  and $RPM_{1}$. Rotation at $RPM_{0}$  results in a vibration frequency of $f_{0}$ (a logical '0'), while rotation at $RPM_{1}$  results in a vibration frequency of $f_{1}$ (a logical '1'). In our modulation we use the return to zero technique in which the signal drops to zero frequency ($RPM_{base}$) between each bit. The operation of the AiR-ViBer FSK modulator is outlined in Algorithm \ref{alg:a1}. Table \ref{t5} provides a summary of the parameters of the FSK modulation. Figure \ref{fig:FSK1}  presents the spectrogram of the FSK modulation, where $RPM_{0}$ = 1300 and $RPM_{1}$ = 2600. In this case the sequence '10101010' has been transmitted.

\begin{table}[!h]
	\centering
	\caption{AiR-ViBeR B-FSK modulation parameters}
	\label{t5}
	\begin{tabular}{llll}
		\hline
		RPM & Carrier Freq. & Duration & Symbol \\ \hline
		$RPM_{0}$   & $f_{0}$            & $T_{0}$        & '0'    \\
		$RPM_{1}$  & $f_{1}$            & $T_{1}$        & '1'    \\
		$RPM_{base}$  & $f_{base}$            & $T_{base}$        & -
		\\ \hline
	\end{tabular}
\end{table}

\begin{algorithm} 
	\caption{modulateASK (fan, $RPM_{0}$, $RPM_{1}$, bits, bitDuration)} 
	\begin{algorithmic}[1] 
		\State $ bitStart = getCurrentTime() $
		\For {$ bit\ in\ bits $}
		\If { $ bit == 1 $ }
		\State $ fan.setRPM(RPM_{1}) $
		\Else
		\State $ fan.setRPM(RPM_{0}) $
		\EndIf
		\\
		\State $ bitStart\ +=\ bitDuration $
		\State $ sleep(bitStart - getCurrentTime())$
		\EndFor
		
	\end{algorithmic}
	\label{alg:aa2}	 
\end{algorithm}

\subsection{Amplitude-shift keying (ASK)}
\begin{figure}
	\centering
	\includegraphics[width=0.8\linewidth]{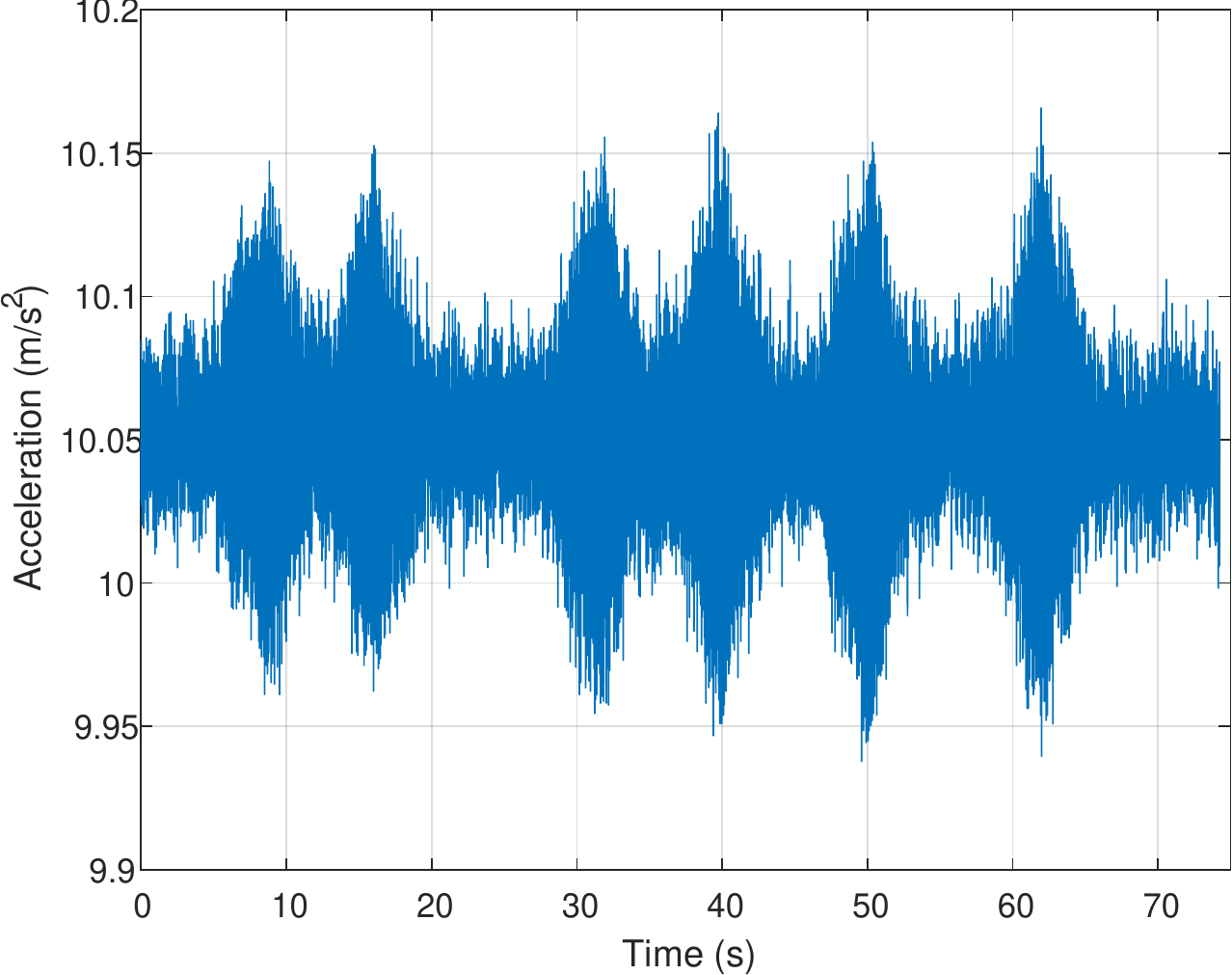}
	\caption{ASK modulation measured 10cm from the vibrating computer at $RPM_{0}$=2000 and $RPM_{1}$=2600}
	\label{fig:ASK1}
\end{figure}

\begin{figure}
	\centering
	\includegraphics[width=0.8\linewidth]{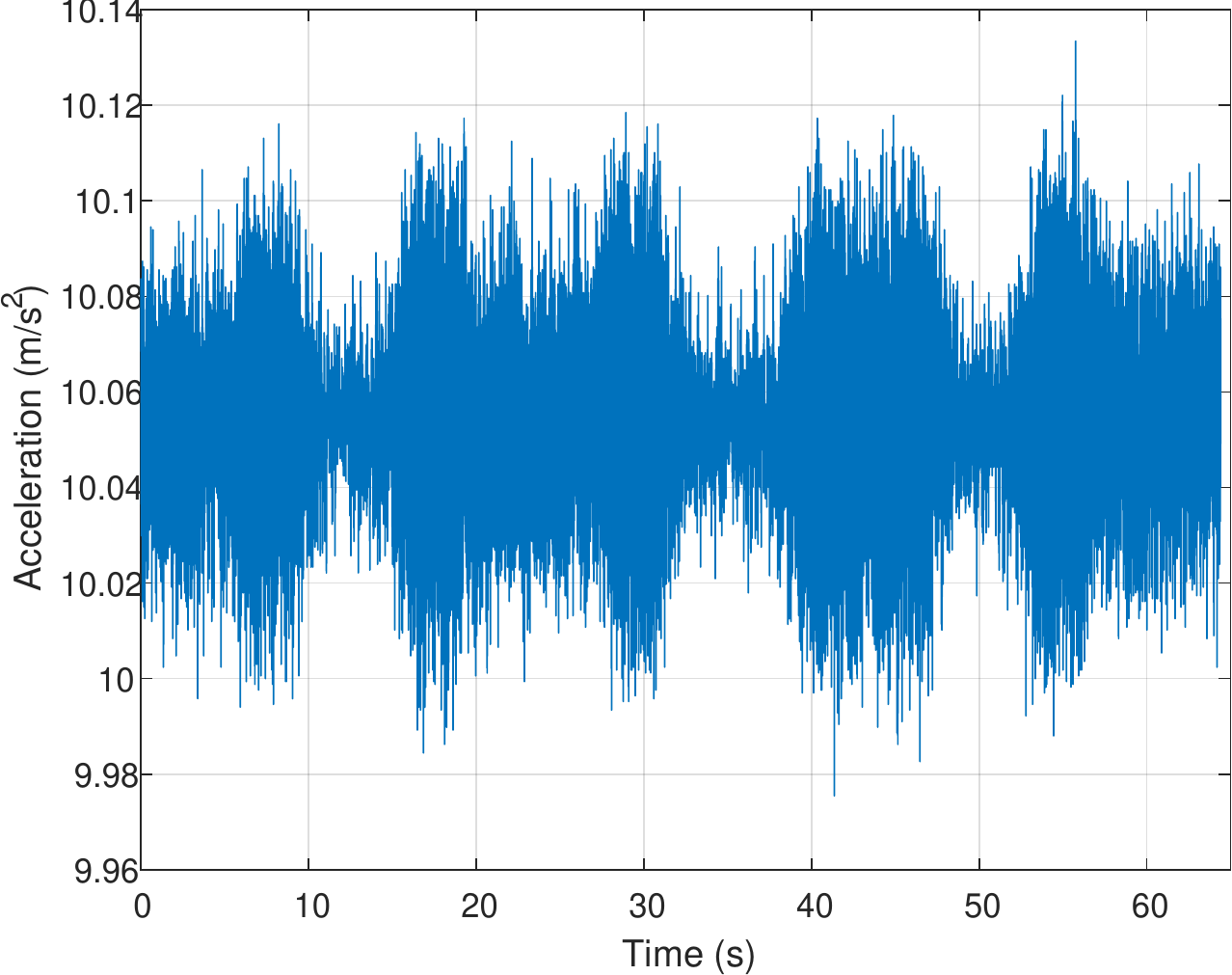}
	\caption{ASK modulation measured 100cm from the vibrating computer at $RPM_{0}$=2000 and $RPM_{1}$=2600}
	\label{fig:ASK2}
\end{figure}

In the amplitude-shift keying modulation we assign distinct amplitude levels of the carrier to represent distinct values of binary data. We use the binary version of ASK (B-ASK), in which two distinct amplitudes, $A_{0}$ and $A_{1}$, represent logical '0' and '1,' respectively. We control the amplitudes by rotating the fan at two different speeds, $RPM_{0}$ and $RPM_{1}$, each at a time period of $T$. The change in the amplitude is caused due to the self resonant frequency of the surface (e.g., the table). When the vibrational frequency meets the self resonant frequency, the amplitude levels increase. 
The operation of the AiR-ViBeR ASK modulator is described in Algorithm \ref{alg:aa2}. The transition between $RPM_{0}$ and $RPM_{1}$ causes a stronger vibration signal. Figures \ref{fig:ASK1} and \ref{fig:ASK2} present the ASK modulation as received from distances of 10cm and 100cm from the transmitting computer, respectively. In these cases the transition between 2000RPM and 2600RPM causes an increase in the vibrations.

\subsection{Demodulation}
We implemented a receiver as an app for the Android OS based on the FSK modulation. The AiR-ViBeR receiver app samples the accelerometer sensor and performs the signal processing. The signals were recorded with a Samsung Galaxy S10 smartphone, running the Android OS version 9.0 "Pie".
The operation of the AiR-ViBeR FSK demodulator app is described in Algorithm \ref{alg:a3}.

\begin{figure}
	\centering
	\includegraphics[width=0.4\linewidth]{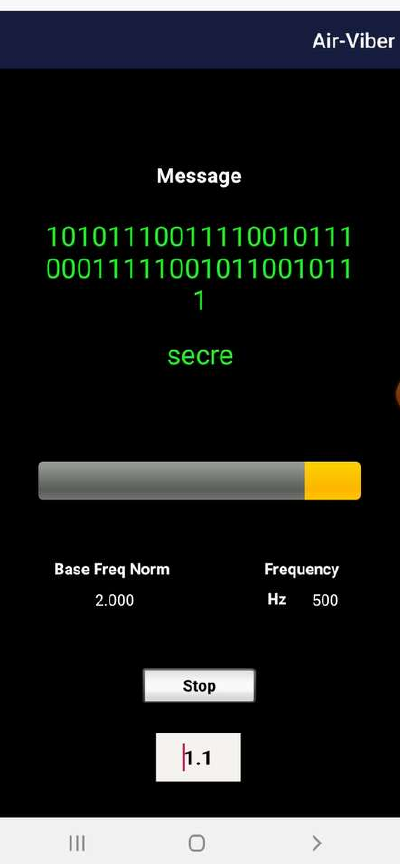}
	\caption{The AiR-ViBeR app while receiving the "secret" word (42 bit) exfiltrated from an air-gapped computer via fan-generated vibrations.}
	\label{fig:app}
\end{figure}

The AiR-ViBeR app (\ref{fig:app}) samples and logs the accelerometer sensor at a frequency of 500Hz. The main function runs in a separate thread. It is responsible for data sampling, signal processing and data demodulation. In our case we sampled the accelerometer sensor using the Android  sensors API. We used the type TYPE\_ACCELEROMETER to sample the acceleration levels.
It then performs a fast Furrier transform (FFT) to the two spectrum in order to measure the power at frequencies $F_0$ and $F_1$. The value of the current bit is determined by the amplitude in these frequencies.

\begin{algorithm} 

	\caption{demodulate(sampleRate, fftSize, noverlap, bitTime, $F_{0}$, $F_{1}$)} 
	\begin{algorithmic}[1] 
		
		\State $ mSensorManager \gets (SensorManager) getSystemService(Context.SENSOR\_SERVICE) $
		\State $ mAccelerometer \gets mSensorManager.getDefaultSensor(Sensor.TYPE\_ACCELEROMETER) $
		\State $ mSensorManager.registerListener(this, mAccelerometer, SensorManager.SENSOR\_DELAY\_FASTEST) $
		\\
		\State $ onSensorChanged(SensorEvent\ event)\ \{ $
		\\
		\State $ enabled \gets False $
		\State $ indexF0 \gets fftSize*F0/sampleRate $
		\State $ indexF1 \gets fftSize*F1/sampleRate $
		\State $ samplesPerBit \gets sampleRate*bitTime/(fftSize - noverlap) $
		
		\State $ magnitude \gets getVectorMagnitude(event) $
		\State $ buffer.append(magnitude)$
		\If{$ buffer.size() == fftSize$}
		\State $ fftWindow \gets fft(buffer) $
		\State $ buffer.removeRange(0, fftSize - noverlap)$
		\\
		\State $ amplitudeF0 \gets abs(fftWindow[indexF0]) $
		\State $ amplitudeF1 \gets abs(fftWindow[indexF1]) $
		\\
		\If {$amplitudeF0 > amplitudeF1$}
		\State $samples.append(0)$
		\Else
		\State $samples.append(1)$
		\EndIf
		\\
		\If {$not\ enabled$}      
		\State enabled = $detectEnable(samples, samplesPerBit)$
		\EndIf
		\\
		\While {$enabled\ \&\ samples.size() >= samplesPerBit$}      
		\If {$2 * sum(samples) >= samplesPerBit$}
		\State $bits.append(1)$
		\Else
		\State $bits.append(0)$
		\EndIf
		\State $samples.removeRange(0, samplesPerBit)$
		\EndWhile
		
		\EndIf
		\State $return\ bits$
		\\
		\State $ \}$
		
	\end{algorithmic}
	\label{alg:a3}
\end{algorithm}

\subsection{Bit-framing}
We transmit the data in small packets composed of a preamble, a payload, and a parity bit.  

\begin{figure}
	\centering
	\includegraphics[width=1.0\linewidth]{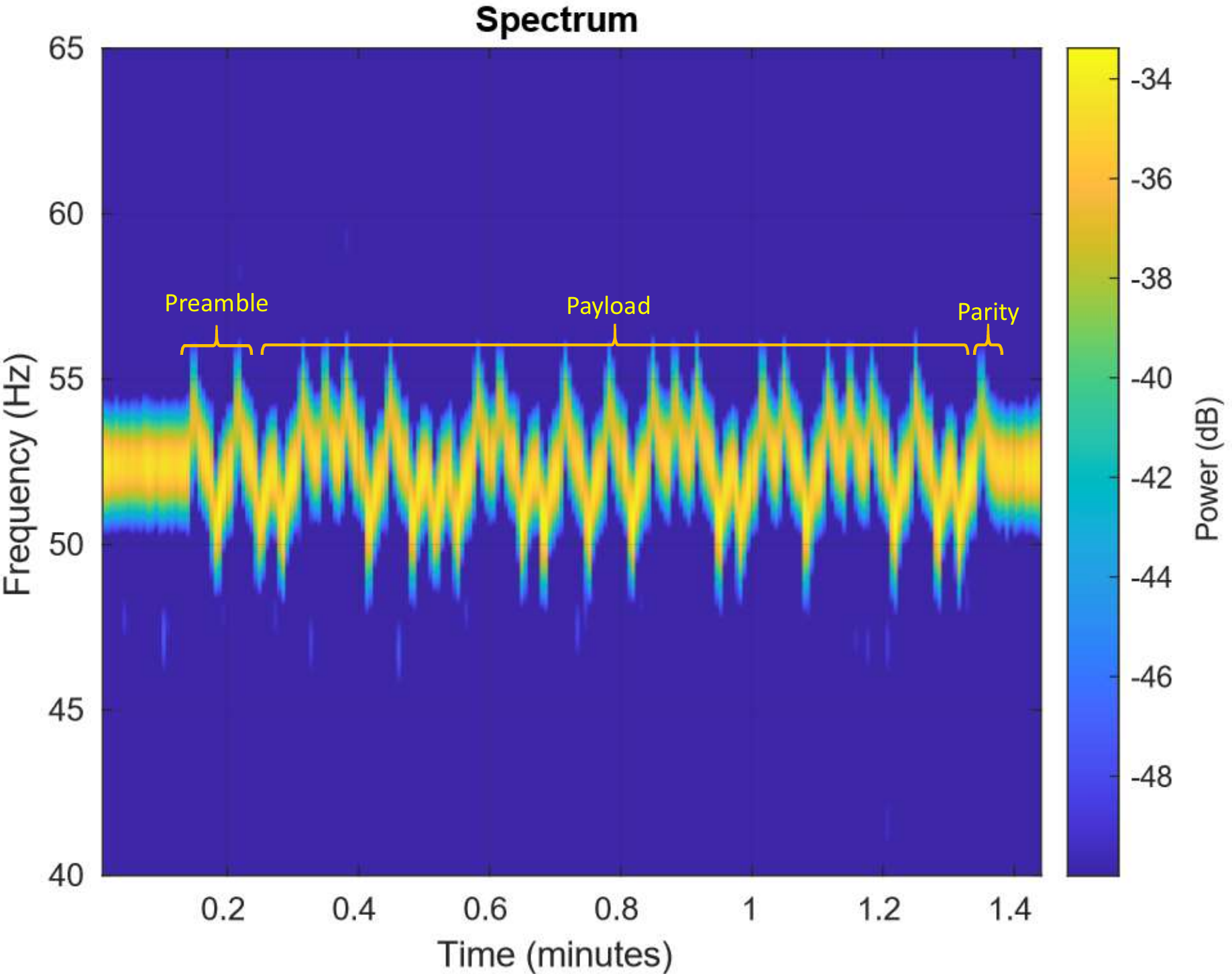}
	\caption{the whole frame as received by a smartphone located on the table.}
	\label{fig:frame}
\end{figure}

\begin{itemize}
	\item	\textbf{Preamble.} A preamble header is transmitted at the beginning of every packet. It consists of a sequence of eight alternating bits ('1010') which helps the receiver determine the carrier wave frequency and amplitude. In addition, the preamble allows the receiver to detect the beginning of a transmission. 
	Note that in our covert channel the amplitude of the carrier wave is unknown to the receiver in advance, and it mainly depends on the vibrating surface and the distance between the transmitter and the receiver. These parameters are synchronized with the receiver during the preamble.  
	\item \textbf{Payload.} The payload is the raw data to be transmitted. In our case, we arbitrarily choose 32 bits as the payload size. 
	\item \textbf{Parity bit.} For error detection we add a parity bit at hte end of each frame. In the more advanced frame an eight bit CRC (cyclic redundancy check) is added to the end of the frame. The receiver calculates the parity or CRC for the received payload, and if it differs from the received parity/CRC, an error is detected. The spectrogram in Figure \ref{fig:frame} shows the full frame (preamble + payload + parity) as received by a smartphone on the table.
\end{itemize}

\section{Evaluation}
Our measurement setup consists of desktop computers for transmission and smartphones installed with the AiR-ViBeR app for the reception ]. 

We used the off-the-shelf desktop PCs listed in Table \ref{tab:desktops}. During the tests, we control the fan speed via the Raspberry Pi as described in Section \ref{sec:imp}. For the measurements we used the Samsung Galaxy S10 smartphone with the Android OS version 9.0 "Pie" installed; accelerometer model STM LSM6DSO is integrated in this smartphone. 

It is important to note that
the vibrational covert channel is highly dependent on the specific environment, structure and material of the surface and the location of the sender and receiver. Providing a comprehensive analytic model for this covert channel in various scenarios is beyond the scope of this paper and left for future work in this field.

\begin{table*}[]
	\centering
	\caption{The desktop workstations used for the evaluation}
	\label{tab:desktops}
\begin{tabular}{@{}lllll@{}}
	\toprule
	\#            & Case              & CPU                             & Board                      & Max RPM \\ \midrule
	Workstation-0 & OptiPlex 9020     & Intel Core i7-4790  CPU 3.60GHz       & DELL 0N4YC8                & 2600    \\
	Workstation-1 & Infinity          & Intel Core i7-4790 CPU 3.60GHz  & Gigabyte GA-H87M-D3H board & 3260    \\
	Workstation-2 & Lenovo ThinCentre & Intel Core i5v 2400 CPU 3.60GHz & Lenovo board               & 3260    \\ \bottomrule
\end{tabular}
\end{table*}

\subsection{Vibration surface}
\begin{figure*}
	\centering
	\includegraphics[width=1.0\linewidth]{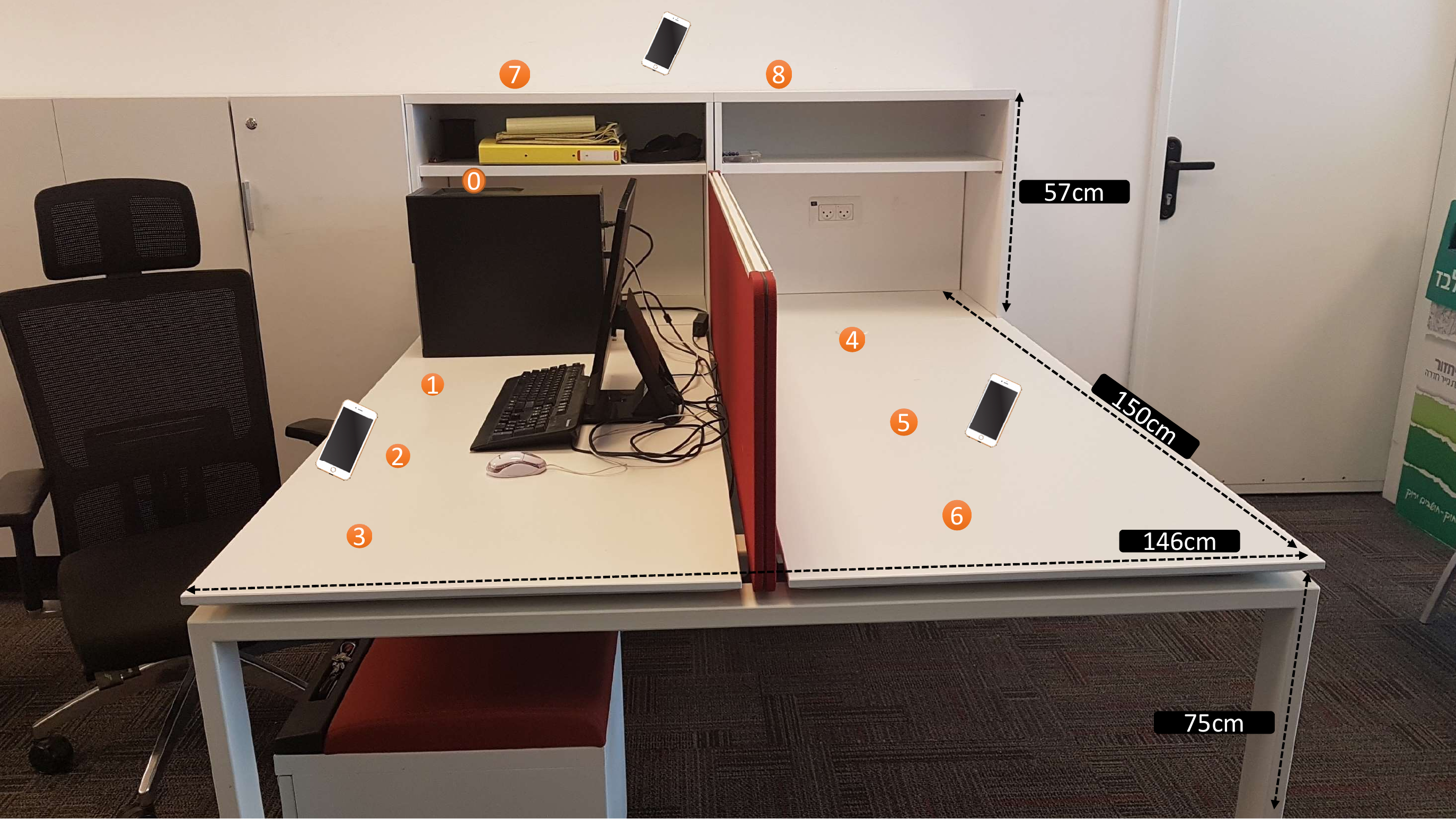}
	\caption{Measurement points on the table}
	\label{fig:tables}
\end{figure*}

We measured the vibration signal as received by the smartphone located in various places on a typical workplace table. The desk shown in Figure \ref{fig:tables} consists of two adjacent tables (150x70cm each) and two cabins. During the test, we transmitted chirp (sweep) signals from the Workstation-0 and sample them with a smartphone at nine different locations on the two tables. We used MATLAB to compute the SNR at a frequency of 43Hz (2580RPM). The results are presented in Table \ref{tab:SNRA}. As can be seen, the PC-generated vibrations can be sensed all over the two tables (points 0-6), and at one of the adjacent cabins as well (point 7).

\begin{table*}[]
	\centering
	\caption{Signal measurements for different locations}
	\label{tab:SNRA}
	\begin{tabular}{@{}llllllllll@{}}
		\toprule
		Location & 0        & 1        & 2        & 3        & 4       & 5        & 6        & 7         & 8      \\ \midrule
		SNR      & 45.02 dB & 21.68 dB & 29.12 dB & 16.81 dB & 22.1 dB & 11.38 dB & 21.43 dB & no clear signal & 7.88dB \\ \bottomrule
	\end{tabular}
\end{table*}

\subsection{Signal-to-noise ratio (SNR)}
\begin{figure}
	\begin{minipage}[b]{0.23\textwidth}
		\includegraphics[width=\textwidth]{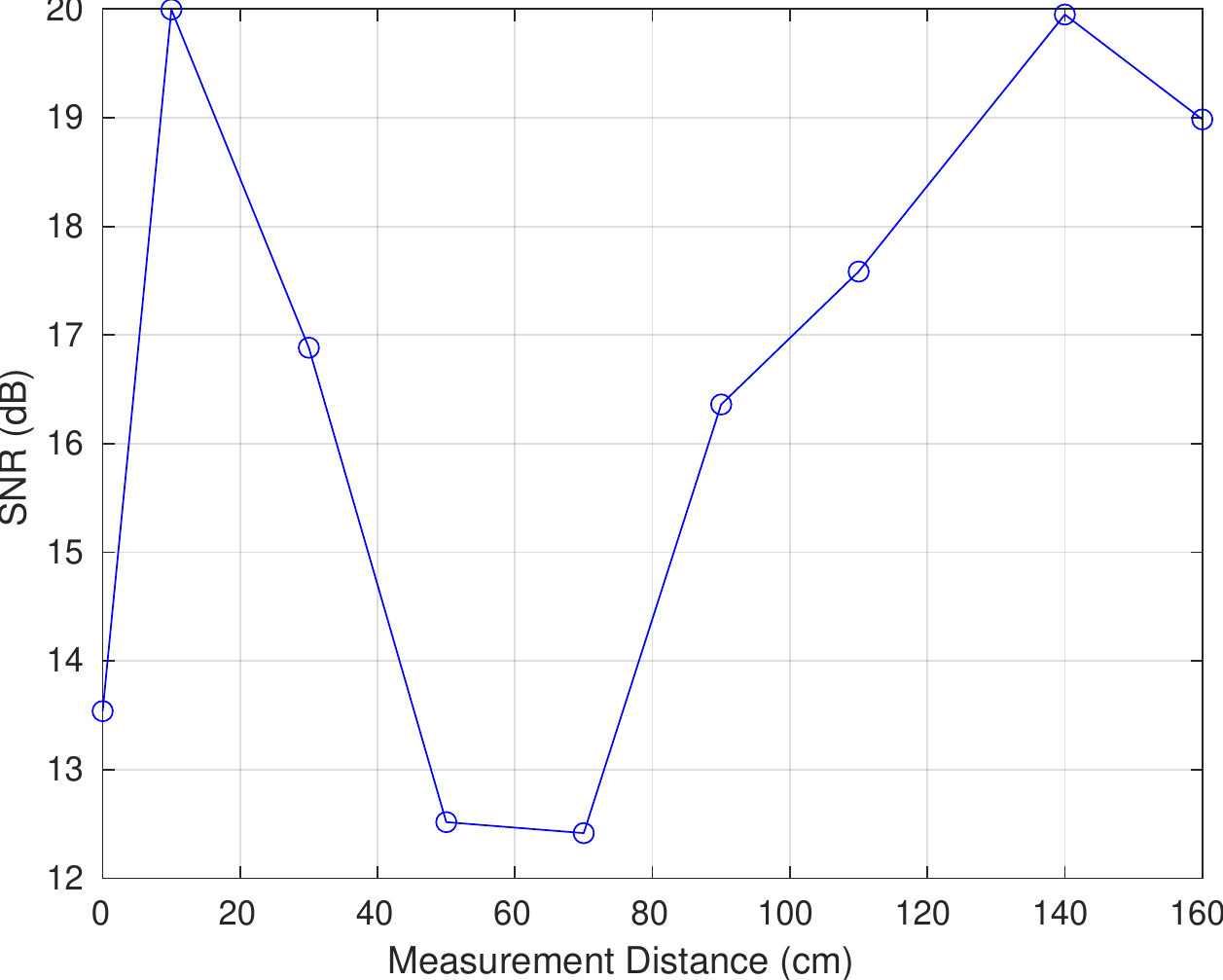}
		\caption{SNR of the transmission from Workstation-1}
		\label{fig:snr1}
	\end{minipage}
	\begin{minipage}[b]{0.23\textwidth}
		\includegraphics[width=\textwidth]{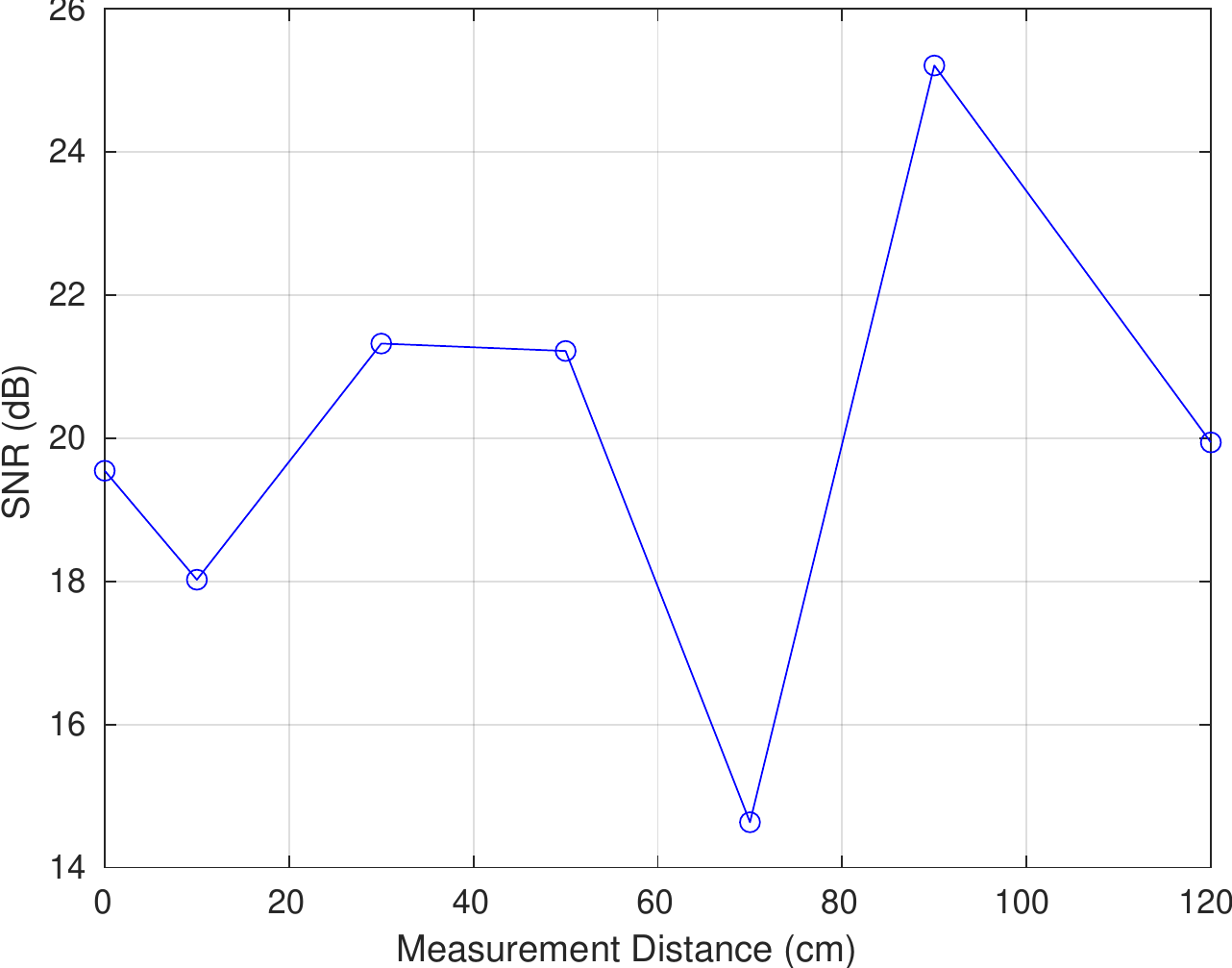}
		\caption{SNR of the transmission from Workstation-2}
		\label{fig:snr2}
	\end{minipage}
\end{figure}

We measured the signal-to-noise (SNR) ratio for the
covert vibrational channel at various distances. During the experiments, the desktop computers and smartphone receiver were located on a flat lab table (with dimensions of 200X100cm). Figures \ref{fig:snr1} and \ref{fig:snr2} present the SNR values of Workstation-1 and Workstation-2, respectively. As can be seen, the SNR values varying depending on the exact position of the receiver on the table. We observed that the SNR may be better at certain table locations, e.g., toward the edge of the table. As noted before, the quality of the vibrational covert channel is highly dependent on the specific environment, structure, and material of the surface and the location of the sender and receiver. These SNR values reflects the quality of the signal and background noise using this specific setup.

\subsection{Bit error rate (BER)}
We measured the bit error rates (BER) for receiver locations. In these experiments we repeatedly transmitted sequences of 37 random bits in the structured packet, decoded them, and compared
the results with the original data. We chose to use FSK modulation, since it is faster and more robust for the purpose of data exfiltration than ASK modulation. We set the $T_{base}$ to 1.5sec and the $T_{0}$ and $T_{1}$ to 0.5sec. The channel properties are summarized in Table \ref{tab:prop}.

\begin{table*}[]
	\centering
	\caption{Bit error rates}
	\label{tab:BER}
	\begin{tabular}{@{}llllllllll@{}}
		\toprule
		\#            & On case & 10cm & 30cm & 50cm & 70cm & 90cm & 110cm & 140cm & 160cm \\ \midrule
		Workstation-1 & 0     & 0  & 0  & 0  & 0  & 0  & 8.1  & 0   & 2.7  \\
		Workstation-2 & 0     & 0  & 5.6 & 0  & 0  & 0  & 0   & -     & -     \\ \bottomrule
	\end{tabular}
\end{table*}

Table {\ref{tab:BER} presents the BER measurements for 
Workstation-1 and Workstation-2. In this case the base frequency was set to 3030 RPM, and the high and low frequencies ('1' and '0') were set to 3260 RPM, and 2600 RPM, respectively. 

\begin{table}[h!]
	\centering
	\caption{Channel properties}
	\label{tab:prop}
	\begin{tabular}{@{}llll@{}}
		\toprule
		& Base     & High     & Low      \\ \midrule
		Frequency & 3030 RPM & 3260 RPM & 2600 RPM \\
		Time      & 1.5 sec  & 0.5 sec  & 0.5 sec  \\ \bottomrule
	\end{tabular}
\end{table}


As can be seen, with Workstation-1 we maintained a BER of 0\% up for a distance up to 140cm from the receiver, and 2.7\% for a distance up to 160cm. With Workstation-2 we maintained BER of 0\% for a distance up to 120cm from the receiver. Practically, this implies that BER close to 0\% can be maintained at distances of 120-150cm in our setup.

\subsection{CPU fans}
We also tested the ability of the CPU fan to generate vibrations for the purpose of exfiltration. The results show that the vibration signals can only be received by a smartphone located on the workstation case or very close to the workstation (up to 5cm away). Figure \ref{fig:cpufac} presents the spectrogram of a sweep signal generated from a CPU fan of a workstation. The signal is sampled by a smartphone placed on the workstation case. Our measurements show that when the receiver is placed on top of the workstation we can obtain a signal with an SNR of 15.15dB. 

\begin{figure}
	\centering
	\includegraphics[width=0.8\linewidth]{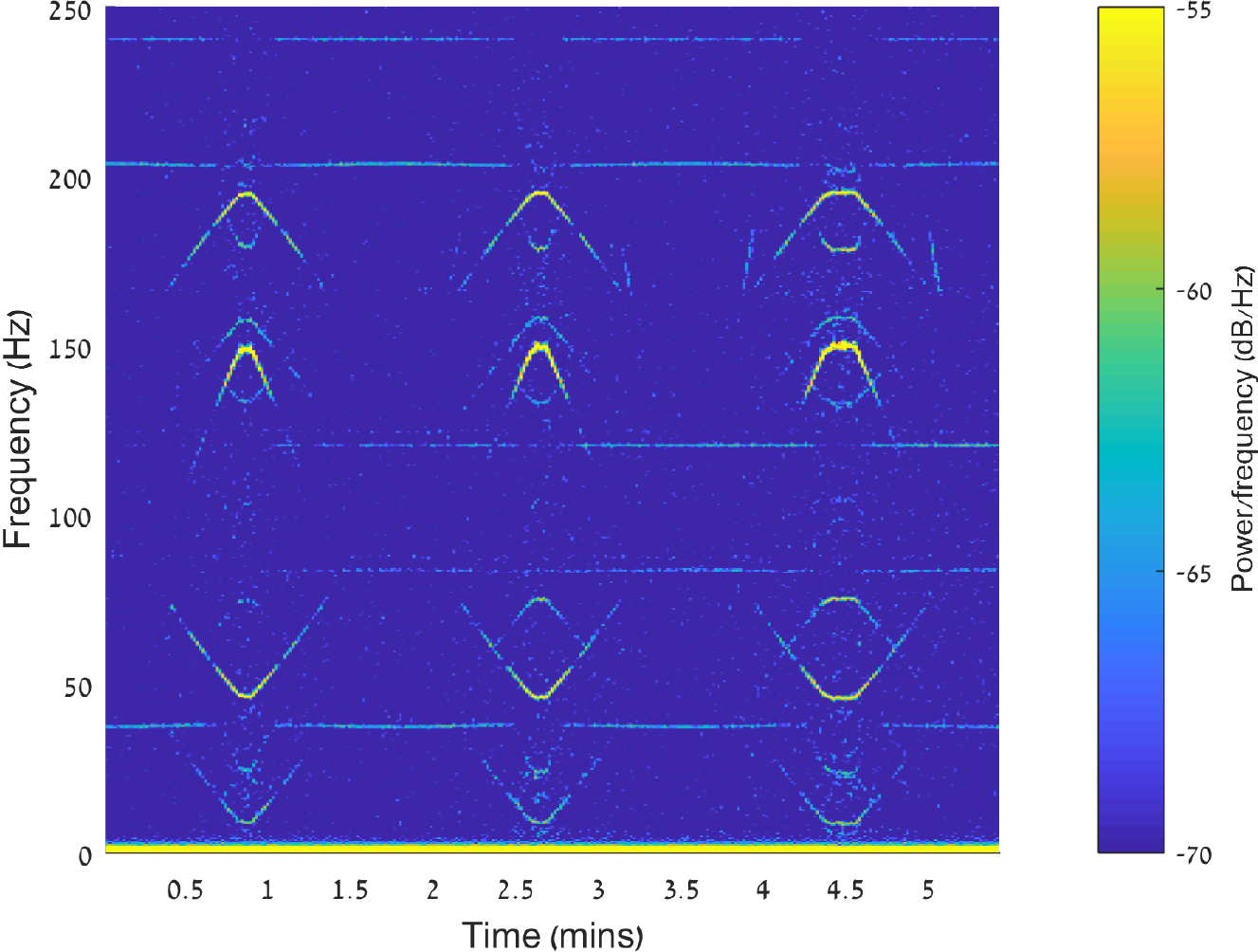}
	\caption{Vibrations generated by a sweep signal of the CPU fan.}
	\label{fig:cpufac}
\end{figure}

\section{Countermeasures}
Defensive countermeasures for general TEMPEST threats
can serve as a source of inspiration for the implementation of countermeasures for the thermal channel. The prevailing standards are aimed at limiting the level of detectable
information-bearing RF signals over a given open-air distance, or its equivalent, when insulating materials are used.
Practically, certified equipment is classified by 'zones' that
refer to the perimeter that needs to be controlled to prevent
signal reception. As a countermeasure against vibration based attacks, the "zones" approach should be used to define
the physical distances required by potential vibrating and
 sensing components connected to different networks. In
some cases, keeping minimal distances between computers and smartphones is not always practical in terms of space and administration
overhead. One
solution may be to place an accelerometer sensor on computers
that contain sensitive information in
order to detect anomalous vibrations. The
logs of these monitoring systems can be used to
detect covert communication attempts if they are made available to
an aware analyst or appropriate policies are put in place.

Software-based countermeasures include the use of endpoint protection to detect code which interferes with the fan control API or accessing the fan controlling bus (e.g., ACPI and SMBus). In a typical system, no process should access the fan control. Such a fan access monitor should be implemented as a low-level driver in the kernel. However, software countermeasures have been shown to be porous \cite{blunden2012rootkit} as an attacker can bypass them using rootkits and other evasion techniques. 

It is also possible to jam the communication channel by interrupting or masking the original transmissions. In the \textit{internal jamming} approach, a dedicated process that changes the fan speed at random times and RPMs is used. However, such a process can be evaded by user level malware unless it is implemented as a kernel driver. Note that even if such a jammer is implemented in the kernel it can be disabled or evaded by kernel rootkits. Pseudo code of such a jammer is shown in Algorithm \ref{alg:a4}. A jammer thread reads the fan speed and changes it periodically at random thresholds. The effect of such a jammer on the communication channel is shown in Figure \ref{fig:jampower}. The figure shows the spectrogram and the power spectral density graph of a jammed communication channel. In this case, we tried to transmit the sequence of bits used in the BER measurements. With the presence of the jammer, we measured BER above 30\%.   As can be seen, the interruptions add constant vibrational noise to the spectrum which masks the original signals.

\begin{algorithm} 

	\caption{jammer (fan, threshold, duration timeInterval)} 
	\begin{algorithmic}[1] 
		\label{alg:4}	
		\State $ currentFanRPM = getFanRPM()$
		\State $ delta = random (-threshold..threshold) $
		
		\State $ fan.setRPM(currentFanRPM + delta) $
		\State $ sleep(duration)$
		\State $ fan.setRPM(currentFanRPM) $
		
		\State $ sleep(timeInterval)$
		
	\end{algorithmic}
	\label{alg:a4}
\end{algorithm}

\begin{figure}
	\centering
	\includegraphics[width=1.0\linewidth]{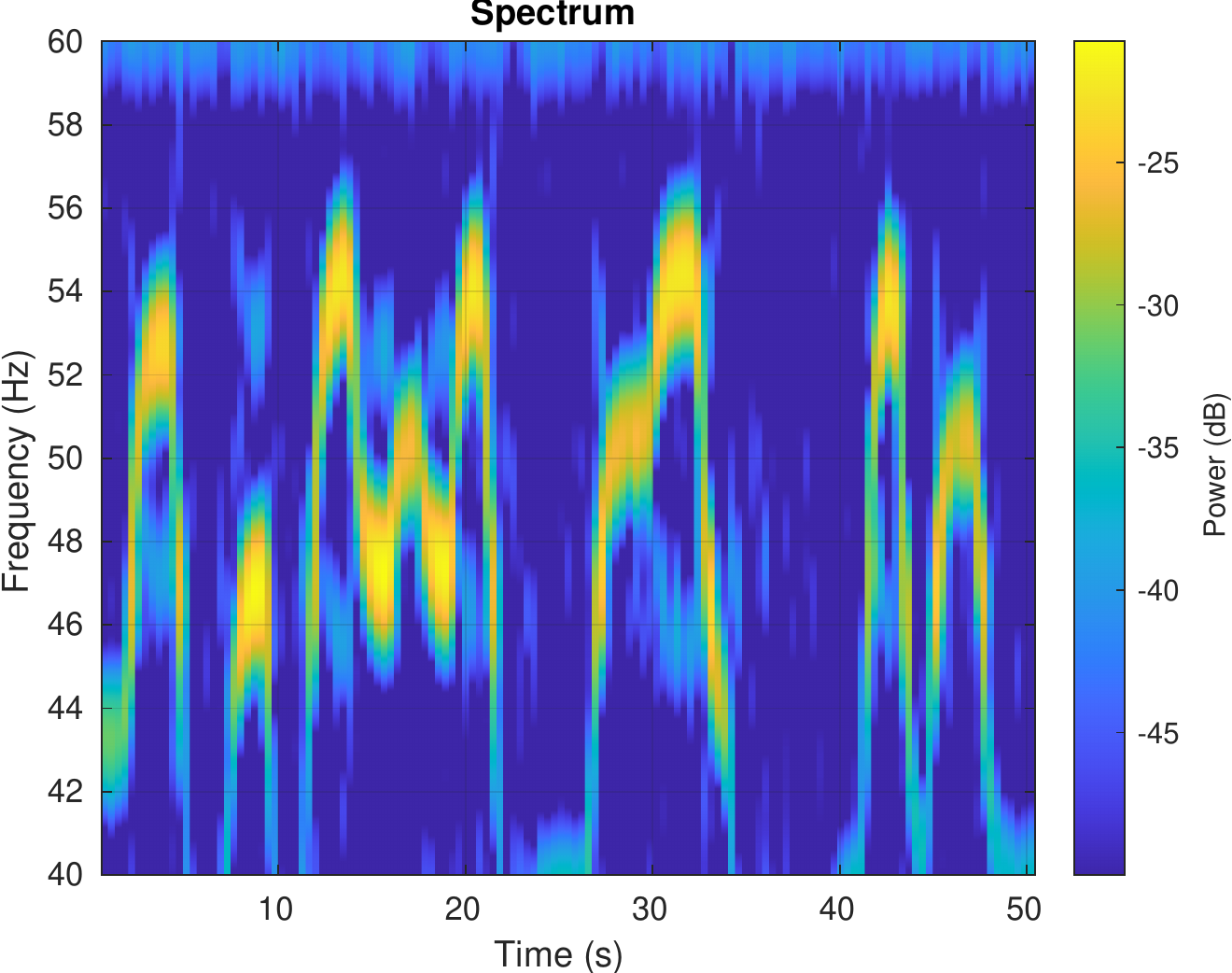}
	\includegraphics[width=1.0\linewidth]{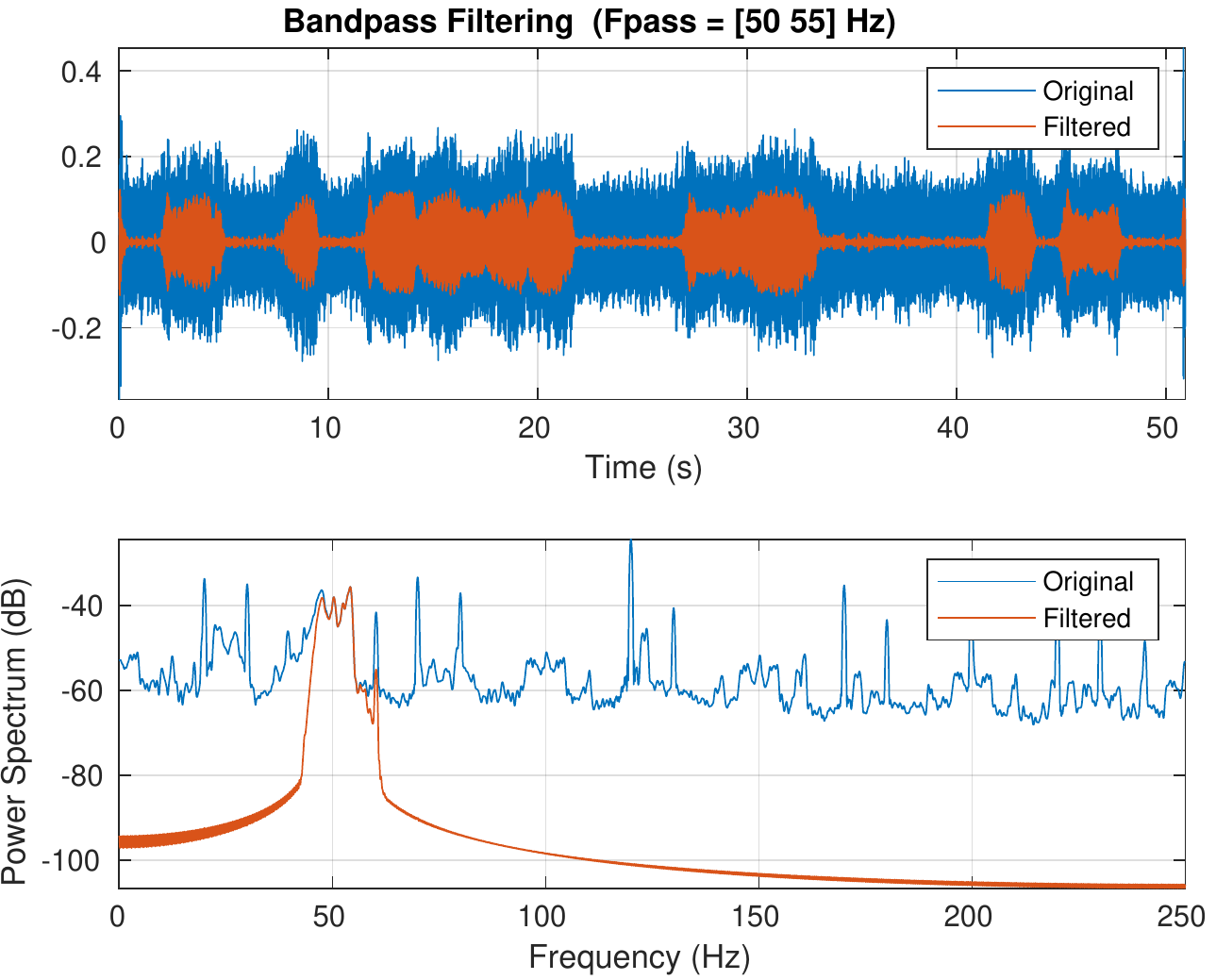}
	\caption{The spectrogram and power spectral density graph of a jammed communication channel}
	\label{fig:jampower}
\end{figure}

In the \textit{external jamming} approach, a dedicated component that generates random (unnoticeable) vibrations is attached to the computer. Such a solution is considered to be trusted in term of security since it can not be accessed by malicious code run on the computer. However this type of solution is impractical for deployment on every computer due to the maintenance it requires (e.g., power source).  

Physical isolation in which the computer chassis is built with special vibration resistance computer cases or low-vibration fans is also an option for limiting the attack \cite{tracy1993computer}. Finally, replacing the original computer fans with  a water cooling system is possible \cite{li2005powerful}. However, such a solution is costly and impractical on a large-scale.

 Table \ref{tab:cntr} contains the list of defensive approaches.

\begin{table}[]
	\centering
	\caption{Countermeasures}
	\label{tab:cntr}
	\begin{tabular}{@{}lll@{}}
		\toprule
		Countermeasure                  & Type       & Cons                  \\ \midrule
		'Zones' approach                & Prevention & Maintenance          \\
		Water cooling systems           & Prevention & Cost                  \\
		Vibration resistance cases and fans & Prevention &  Cost and maintenance                \\
		
		Vibration monitoring         & Detection  & Can be evaded         \\
		Vibration jamming         & Jamming  & Can be evaded         \\
		Generate random vibrations & Jamming    & Cost and maintenance \\ \bottomrule
	\end{tabular}
\end{table}

\section{Conclusion}
In this paper, we introduce a new type of vibration-based covert channel.
We show that malware running on a computer can generate vibrations via the chassis fan in a controlled manner. 
The unnoticeable vibrations can be sensed by a smartphone located on a table using its accelerometer. Binary data can be modulated on top of the vibrations; then it can be decoded on the smartphone and sent to the attacker via the Internet. We discuss the implementation of the transmitter and receiver and present a transmission protocol. We evaluate the communication channel, and in a typical workplace scenario and show that data can be exfiltrated at a speed of half a bit per second via the covert vibrations. Finally, we discuss a set of defensive countermeasures.

\balance
\bibliographystyle{ieeetran}
\bibliography{AirGap,vib}
\end{document}